\documentclass[epj]{webofc}
\usepackage[utf8]{inputenc}
\usepackage[varg]{txfonts}   
\usepackage{booktabs}
\usepackage{xcolor}
\definecolor{darkred}{rgb}{0.4,0.0,0.0}
\definecolor{darkgreen}{rgb}{0.0,0.4,0.0}
\definecolor{darkblue}{rgb}{0.0,0.0,0.4}
\usepackage[bookmarks,linktocpage,colorlinks,
    linkcolor = darkred,
    urlcolor  = darkblue,
    citecolor = darkgreen]{hyperref}
\usepackage{textcomp}
\usepackage{slashed}
\usepackage[FIGBOTCAP,TABBOTCAP,tight]{subfigure}
\usepackage{comment}
\usepackage{indentfirst}
\usepackage{mathrsfs}
\usepackage{subfigure}
\allowdisplaybreaks

\graphicspath{{./figs/}}
\wocname{EPJ Web of Conferences}
\woctitle{Lattice2017}
\renewcommand{\mathcal}{\mathscr}

\newcommand{\epsK}{\varepsilon_{K}}

\begin{document}
%
\selectlanguage{english}
\title{
Improvement of heavy-heavy current for calculation of 
\\
$\bar{B}\to D^{(*)}\ell\bar{\nu}$ form factors using 
Oktay-Kronfeld heavy quarks 
}
\author{
\firstname{Jon~A.} \lastname{Bailey}\inst{1} \and
\firstname{Yong-Chull} \lastname{Jang}\inst{2}\inst{3} \and
\firstname{Weonjong} \lastname{Lee}\inst{1}\fnsep\thanks{\email{wlee@snu.ac.kr}}
\and
\firstname{Jaehoon} \lastname{Leem}\inst{1}\fnsep\thanks{Speaker} \and
\lastname{(LANL-SWME Collaboration)}
}
\institute{%
  Lattice Gauge Theory Research Center, CTP,
  Department of Physics and Astronomy, \\
  Seoul National University, Seoul 08826, South Korea
  \and
  Los Alamos National Laboratory,
  Theoretical Division T-2,
  Los Alamos, New Mexico 87545, USA
  \and
  Brookhaven National Laboratory,
  Department of Physics, Upton, New York 11973, USA
}
\abstract{The CKM matrix element $|V_{cb}|$ can be extracted by
  combining data from experiments with lattice QCD results for the
  semileptonic form factors for the $\bar{B}\to D^{(*)} \ell
  \bar{\nu}$ decays.  The Oktay-Kronfeld (OK) action was designed to
  reduce heavy-quark discretization errors to below $1\%$, or through
  $\mathcal{O}(\lambda^3)$ in HQET power counting.  Here we describe
  recent progress on bottom-to-charm currents improved to the same
  order in HQET as the OK action, and correct formerly reported
  results of our matching calculations, in which the operator basis
  was incomplete. }
\maketitle
\section{Introduction}
\label{sec:intro}
The Cabibbo-Kobayashi-Maskawa (CKM) matrix element $|V_{cb}|$
normalizes the unitarity triangle, and the uncertainty in $|V_{cb}|$
propagates to many quark-flavor observables such as $\epsK$ \cite{
  Lee:2017proc}, limiting the precision of constraints on the CKM
matrix.
Accordingly, determinations of $|V_{cb}|$ with greater precision are
essential to improve these constraints and understand the mechanism
underlying the dynamics and CP violation of the quark-flavor sector.

$|V_{cb}|$ can be extracted from the exclusive semileptonic decays
$\bar{B}\to D \ell \bar{\nu}$ and $\bar{B}\to D^* \ell \bar{\nu}$.
The approach involves combining the branching fractions from
experiments with the form factors calculated on the lattice~\cite{
  Bailey:2014tva, Lattice:2015rga, Na:2015kha}.
The calculations of the form factors by the authors of Refs.~\cite{
  Bailey:2014tva, Lattice:2015rga} have so far relied on the Wilson
clover action \cite{ Sheikholeslami:1985ij} interpreted
nonrelativistically \textit{via} HQET \cite{ ElKhadra:1996mp,
  Kronfeld:2000ck} to control the discretization errors of the charm
and bottom quarks, with final errors of $2-5\%$ \cite{ Bailey:2014tva,
  Lattice:2015rga}.
%
Given precise lattice inputs for the form factors, the data from
BELLE2 at KEK will allow extractions of $|V_{cb}|$ at the subpercent
level.

The Oktay-Kronfeld (OK) action is an improved Wilson action developed
to reduce the discretization errors of heavy quarks to below $1\%$
even on lattices with spacings as large as $a \approx
0.12\ \mathrm{fm}$~\cite{PhysRevD.78.014504}.
For quark masses large compared with the lattice cutoff, the OK action
possesses heavy-quark symmetry.
To design the OK action, it is convenient to use tools of HQET~\cite{
  Eichten:1989zv} and NRQCD~\cite{ Lepage:1992tx} to quantify the heavy-quark
discretization errors and to tune the action to the continuum
limit \cite{ PhysRevD.78.014504, Kronfeld:2000ck}.
For heavy-light systems, HQET power counting is appropriate, while for
quarkonia, NRQCD provides the appropriate power counting~\cite{
  PhysRevD.78.014504}.
The OK action includes all dimension-five, all dimension-six, and some
dimension-seven bilinears through $\mathcal{O}(\lambda^3)$ in HQET.
Estimates of truncation errors indicate that one-loop matching of
dimension-five bilinears and tree-level matching of the dimension-six
and -seven operators should suffice for the target precision \cite{
  PhysRevD.78.014504}.
Numerical tests of the tree-level matched, tadpole-improved OK action
indicate significant improvement even without one-loop matching of the
dimension-five bilinears~\cite{ Bailey:2017nzm}.

To reduce systematically heavy-quark discretization errors in
calculations of the $\bar{B}\rightarrow D^{(*)} \ell \bar{\nu}$ form
factors, the flavor-changing ($b\to c$) currents must also be improved
through $\mathcal{O}(\lambda^3)$ in HQET.
In addition to the dimension-four improvement term introduced in
Ref.~\cite{ElKhadra:1996mp}, the improved current must include dimension-five
and -six operators.
In Sec.~\ref{sec:current-improvement}, we introduce an improved current in
terms of an improved quark field.
In Sec.~\ref{sec:matching-calculation}, we write down matching conditions and
describe a tree-level matching calculation \textit{via} HQET.
In Sec.~\ref{sec:results}, we discuss the results. 
We add two appendices to describe some technical details.
%

\section{$\mathcal{O}(\lambda^3)$-improved quark field}
\label{sec:current-improvement}

To take advantage of using the OK action in calculating form factors
of $\bar{B}\to D^{(*)}\ell \bar{\nu}$ semileptonic decays, the
improvement of flavor-changing currents should be performed to the
same level, \textit{i.e.}, through $\mathcal{O}(\lambda^3)$ in HQET
power counting.
%
%
%
In case of improvement through $\mathcal{O}(\lambda)$, the current
improvement was performed by introducing an improved quark field
\cite{ElKhadra:1996mp},
\begin{align}
\label{eq:improved-current}
J^{\text{lat}}_\Gamma \equiv \bar{\Psi}_{Ic}\Gamma \Psi_{Ib},
\end{align}
%
where $\Psi_{If}$ is the improved quark field and $\Gamma$ represents
the Dirac spin structure.
%
%

Our basic assumption is that improving quark fields is sufficient for
the improvement of currents at the tree level \cite{ ElKhadra:1996mp}.
To improve the current up to $\mathcal{O}(\lambda^3)$, we extend the
construction of the improved quark fields to higher dimension.
To obtain the results of Ref.~\cite{Bailey:2016wza}, we assumed an
ansatz for the improved quark field motivated by the symmetries of the
theory and the operators appearing in the OK action, which turns out
to be insufficient because two operators were missing from the
operator basis.
These two additional operators are essential to match four quark
matrix elements at $\mathcal{O}(\boldsymbol{p})$ in the external
heavy-quark momentum and at $\mathcal{O}(\lambda^3)$ in HQET.
%

%
%
Considering the Foldy-Wouthuysen-Tani (FWT) transformation on the continuum
quark action, we are led to include two new operators in the ansatz.
To $\mathcal{O}(1/m^3)$, the FWT transformation \cite{ Balk:1993ev}
used to obtain the $\mathcal{O}(1/m^3)$ HQET Lagrangian \cite{
  Manohar:1997qy} from the Dirac action is
\begin{align}
\label{eq:HQET-FWT-1}
Q =
\Bigg[ 1 &- \frac{1}{2m} \boldsymbol{\gamma \cdot D} 
+\frac{1}{8m^2}\boldsymbol{D}^2
+\frac{i}{8m^2}\boldsymbol{\Sigma}\cdot \boldsymbol{B}
+\frac{1}{4m^2}\boldsymbol{\alpha}\cdot \boldsymbol{E}
-\frac{\{\gamma_4D_4,\boldsymbol{\alpha}\cdot \boldsymbol{E}\}}{8m^3}
-\frac{3\{\boldsymbol{\gamma \cdot D},\boldsymbol{D}^2\}}{32m^3}
\nonumber \\
&
-\frac{3\{\boldsymbol{\gamma \cdot D},i\boldsymbol{\Sigma}\cdot \boldsymbol{B}\}}{32m^3}
-\frac{\{\boldsymbol{\gamma}\cdot \boldsymbol{D}, 
\boldsymbol{\alpha}\cdot \boldsymbol{E}\}}{16m^3}
+\frac{[\gamma_4 D_4,\boldsymbol{D}^2]}{16m^3}
+\frac{[\gamma_4 D_4,i\boldsymbol{\Sigma} \cdot \boldsymbol{B}]}{16m^3}
\Bigg]h
+\mathcal{O}(1/m^4),
\end{align}
where $Q$ is the Dirac field, and $h$ is the heavy quark field in HQET.
Accordingly, we consider an ansatz for the improved field that
includes all operators analogous to those in the continuum FWT
transformation of Eq.~\eqref{eq:HQET-FWT-1} as well as those
corresponding to lattice artifacts,
\begin{align}
\label{eq:improved-current-lambda-third}
&\Psi_{I}(x) 
= e^{m_{1}/2}
\bigg[1+d_{1} \boldsymbol{\gamma} \cdot \boldsymbol{D}_{\text{lat}}  
+\frac{1}{2}d_2 \Delta^{(3)} 
+\frac{1}{2}i d_B \boldsymbol{\Sigma}\cdot\boldsymbol{B}_{\text{lat}}
+\frac{1}{2}d_E \boldsymbol{\alpha} \cdot \boldsymbol{E}_{\text{lat}}
+ \frac{1}{6}d_{3} \gamma_i D_{\text{lat},i} \Delta_{i} 
+ d_{EE}\{\gamma_4 D_{\text{lat},4}, \boldsymbol{ \alpha} \cdot \boldsymbol{E}_{\text{lat}}\} 
\nonumber \\
&
+ \frac{1}{2}d_{4}\{\boldsymbol{ \gamma} \cdot \boldsymbol{D}_{\text{lat}},\Delta^{(3)} \}
+ d_{5}\{\boldsymbol{ \gamma} \cdot \boldsymbol{D}_{\text{lat}} ,i \boldsymbol{ \Sigma} \cdot \boldsymbol{B}_{\text{lat}}\}
+ d_{r_E} \{ \boldsymbol{\gamma} \cdot \boldsymbol{D}_{\text{lat}}, \boldsymbol{\alpha} \cdot \boldsymbol{E}_{\text{lat}}\} 
+d_6 [\gamma_4 D_{\text{lat},4} ,\Delta^{(3)} ]
\nonumber \\
&+d_7 [\gamma_4 D_{\text{lat},4}, i\boldsymbol{\Sigma}\cdot \boldsymbol{B}_{\text{lat}} ]
+ d_{z_3} \boldsymbol{\gamma} \cdot (\boldsymbol{D}_{\text{lat}}\times \boldsymbol{B}_{\text{lat}}
+ \boldsymbol{B}_{\text{lat}}\times \boldsymbol{D}_{\text{lat}}) 
+ d_{z_E} \gamma_4(\boldsymbol{D}_{\text{lat}} \cdot \boldsymbol{E}_{\text{lat}}-\boldsymbol{E}_{\text{lat}} \cdot \boldsymbol{D}_{\text{lat}}) 
\bigg] \psi(x),
\end{align}
where we set $a=1$ for convenience.

The terms in Eq.~\eqref{eq:improved-current-lambda-third} with coefficients
$d_3$, $d_{z_3}$, and $d_{z_E}$ have no analogues in Eq.~\eqref{eq:HQET-FWT-1}.
The $d_{3}$ term breaks rotational symmetry and is necessary to remedy
symmetry breaking.
In Eq.~\eqref{eq:HQET-FWT-1}, the $d_{z_3}$ and $d_{z_E}$ terms are
absent, which reflects the fact that they vanish at tree level,
as reported in Ref.~\cite{Bailey:2016wza}.
Comparing Eq.~\eqref{eq:improved-current-lambda-third} with the ansatz
of Ref.~\cite{Bailey:2016wza}, we observe that the terms for $d_6$ and
$d_7$ are new.
These correspond to the last two terms in the FWT
transformation of Eq.~\eqref{eq:HQET-FWT-1}.
In our previous paper \cite{Bailey:2016wza}, our calculation at
$\mathcal{O}(\boldsymbol{p})$ was incomplete, so that we missed the
$d_6$ and $d_7$ terms.
An explicit matching calculation of the lattice and continuum
four-quark matrix elements \textit{via} HQET indicates that the ansatz
with the $d_6$ and $d_7$ terms suffices to match these matrix elements
at tree level through $\mathcal{O}(\lambda^3)$.

\section{Matching calculation}
\label{sec:matching-calculation}
In this section we describe how to determine the coefficients $d_i$ in
the improved quark field.
We match the following four-quark matrix elements at tree level
between the lattice theory and continuum QCD,
\begin{align}
  \langle \ell(\eta_2, p_2) c(\eta^\prime, p^\prime)| \;
  \bar{\Psi}_{Ic} \Gamma \Psi_{Ib}
  \; | b(\eta, p) \ell(\eta_1, p_1)\rangle_\text{Lat}
  \leftrightarrow
  \langle \ell(\eta_2, p_2) c(\eta^\prime, p^\prime)| \;
  \bar{c} \Gamma b
  \; | b(\eta, p) \ell(\eta_1, p_1)\rangle_\text{Con} \,,
  \label{eq:4q-ME}
\end{align}
where $\ell$ represents a light spectator quark, and $c$ and $b$
indicate a charm quark and bottom quark, respectively.
At tree level gluon exchange may occur at either the $b$-quark line or the
$c$-quark line.
The matching conditions for the coefficients in the improved $b$- and
$c$-quark fields factorize at the tree level, and the coefficients in
each field are determined separately.
Hence, we consider only gluon exchange at the $b$-quark line, because
the results for the $c$-quark field are formally identical.
\begin{figure}[tbhp]
  \renewcommand{\subfigcapskip}{-0.95em}
  \vspace*{-5mm}
  \subfigure[One gluon emission from the OK action]{
    \label{fig:1g-act}
    \includegraphics[width=0.40\textwidth]{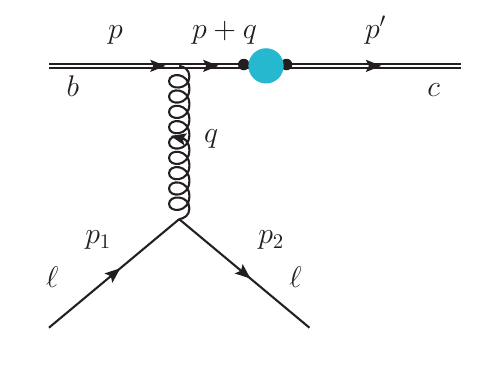}
  }
  \hfill
  \subfigure[One gluon emission from the field improvement terms]{
    \label{fig:1g-imp}
    \includegraphics[width=0.40\textwidth]{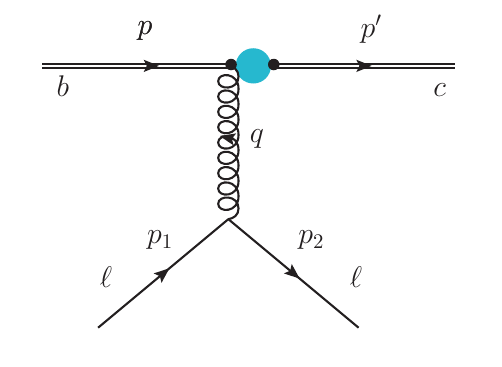}
  }
  \vspace*{-3mm}
  \caption{ The tree level, lattice diagrams with gluon exchange at
    the external $b$-quark line.  The dot without a gluon line in
    \subref{fig:1g-act} diagram represents the zero-gluon vertex, and
    the dot with a gluon line in \subref{fig:1g-imp} diagram
    represents the one-gluon vertex from the improved quark field.}
  \label{fig:latt-diagram}
\end{figure}
%

%
The corresponding lattice Feynman diagrams are given in 
Figure \ref{fig:latt-diagram}.
One gluon exchange diagrams show up in two different ways: The
one-gluon vertex of the action in
Fig.~\ref{fig:latt-diagram}\,\subref{fig:1g-act} can emit a gluon, and
the improvement vertex of the $b$ quark in
Fig.~\ref{fig:latt-diagram}\,\subref{fig:1g-imp} can do it, too.
Comparing these lattice diagrams with the continuum diagrams provides
the following matching conditions for the improved $b$-quark field,
\begin{align}
& n_\mu(q)
\Big[ R^{(0)}_b(p+q)  S^{\text{lat}}(p+q)(-gt^a) \Lambda_\mu(p+q,p)
+(-gt^a)R^{(1)}_{b\mu}(p+q,p)\Big]
\mathcal{N}_b(\boldsymbol{p})u^{\text{lat}}_b(\eta,\boldsymbol{p})
\nonumber \\
& =
S(p+q)
(-gt^a)\gamma_\mu 
\sqrt{\frac{m_b}{E_b}}u_b(\eta,\boldsymbol{p}) \,,
\label{eq:matching-subdiagram}
\end{align}
where the LHS (RHS) represents the lattice (continuum) part.
Here, $q$ is the four-momentum of the gluon, and $\mu$ and $a$ are the
Lorentz and color indices, respectively.
$n_\mu(q)=2\sin(\frac{1}{2}q_\mu)/q_\mu$ is the wave function factor
for the lattice gluon.
$\mathcal{N}_b(p)$ is the lattice spinor normalization factor for the
$b$ quark, which corresponds to $\sqrt{{m_b}/{E_b}}$ in the continuum.
$S_b$ and $S^{\text{lat}}_b$ are the $b$-quark propagator in the continuum 
and on the lattice, respectively.
$\Lambda_\mu$ is the one-gluon vertex of the OK action 
\cite{PhysRevD.78.014504}.

The contributions of the improvement parameters $d_i$ enter through
the zero-gluon vertex $R^{(0)}_b$ and the one-gluon vertex
$R^{(1)}_{b\mu}$ from the improved quark field.
The explicit formulas were given in Ref.~\cite{Bailey:2016wza}.
The results for $R^{(1)}_{b\mu}$ must be corrected for the addition of
the $d_6$ and $d_7$ terms; the results for $R^{(0)}_b$ are unchanged.
Then we have
\begin{align}
R^{(1)}_{bi} &= R^{(1)\text{Ref.~\cite{Bailey:2016wza}}}_{bi} +e^{m_{1,b}}
\big[\sin(p_4+q_4) -\sin p_4 \big]
\Bigg[2 d_6 \gamma_4 \sin(p_i + \frac{1}{2}q_i)
 -i \epsilon_{ijk}d_7 \cos \frac{1}{2}q_i \sin q_j \Sigma_k \gamma_4 
\Bigg],
\\
R^{(1)}_{b4} &= R^{(1)\text{Ref.~\cite{Bailey:2016wza}}}_{b4} +4e^{m_{1,b}}
d_6 \cos (p_4+\frac{1}{2}q_4 )
\sum_{j=1}^3 \Big[ \sin^2 \frac{1}{2}(p_j +q_j)
-\sin^2 \frac{1}{2}p_j \Big].
\end{align}

The spatial momentum of the $b$ quark $\boldsymbol{p}$ and the
four-momentum of the gluon $q$ are of order $\Lambda_{\text{QCD}}$,
small compared to $m_b$ and the lattice cut-off scale $1/a$.
Expanding both sides of Eq.~\eqref{eq:matching-subdiagram} in powers
of $\boldsymbol{p}/m_b$, $q/m_b$, $\boldsymbol{p}a$, and $qa$, all the
terms are organized by powers of $\lambda \sim a \Lambda_{\text{QCD}}
\sim \Lambda_{\text{QCD}}/2m_b$.
Expanding through $\mathcal{O}(\lambda^3)$, both sides of
Eq.~\eqref{eq:matching-subdiagram} can be sorted into terms of the
zero- and one-gluon vertices of the lattice and continuum HQET
Lagrangians and flavor-changing currents~\cite{Kronfeld:2000ck}.
Explicitly, the LHS of Eq.~\eqref{eq:matching-subdiagram} can be
rewritten as follows,
\begin{align}
\label{eq:HQET-description-lattice}
\Bigg[R^{\text{lat},(1)}_{\text{HQ},\mu}(p+q,p) 
+\sum_n R^{\text{lat},(0)}_{\text{HQ}}(p+q)
\bigg(\frac{1}{iP^{\text{lat}}_4}\Lambda^{\text{lat},(0)}_{\text{HQ}}(p+q)\bigg)^n
\frac{1}{iP^{\text{lat}}_4}\Lambda^{\text{lat},(1)}_{\text{HQ},\mu}(p+q,p)
\Bigg]
(-gt^a)u(\eta,0),
\end{align}
where 
$P^{\text{lat}}_4 \equiv p^{\text{lat}}_4 -im_{1,b} +q_4$
is the time component of the residual momentum of the internal $b$ quark.
$\Lambda^{\text{lat},(0)}_{\text{HQ}}$ and
$\Lambda^{\text{lat},(1)}_{\mu,\text{HQ}}$ are the zero- and one-gluon vertices
of the lattice HQET Lagrangian, respectively.
$R^{\text{lat}, (0)}_{\text{HQ}}$ and $R^{\text{lat},
  (1)}_{\text{HQ},\mu}$ are the zero- and one-gluon vertices from the
lattice HQET current, including the correction terms analogous to
those in the FWT transformation of Eq.~\eqref{eq:HQET-FWT-1}.
Expressions for the vertices are given in Appendix~\ref{app:lhqet}.
The external $b$ quark is on shell, 
\begin{align}
p^{\text{lat}}_4 -im_{1,b} = i~\left[ \frac{1}{2m_{2,b}}\boldsymbol{p}^2
-\frac{1}{6}w_{4,b} \sum_i p_i^4- \frac{1}{8m_{4,b}^2} \boldsymbol{p}^4 +\cdots
\right],
\end{align}
where matching the dispersion relation requires $m_{2,b}=m_{4,b} =m_b$ and
$w_{4,b} = 0$~\cite{ElKhadra:1996mp}.  
In terms of the continuum HQET vertices, the form of the RHS of
Eq.~\eqref{eq:matching-subdiagram} is the same as
Eq.~\eqref{eq:HQET-description-lattice}, with all superscripts ``lat''
dropped.

The matching condition Eq.~\eqref{eq:matching-subdiagram} is that the
lattice HQET vertices are equal to those in the continuum as follows,
%
\begin{align}
\Lambda^{\text{lat},(0)}_{\text{HQ}}(p+q,p) &=
\Lambda^{(0)}_{\text{HQ}}(p+q,p),
\qquad
\Lambda^{\text{lat},(1)}_{\text{HQ},\mu}(p+q,p) =
\Lambda^{(1)}_{\text{HQ},\mu}(p+q,p) \label{eq:match-lagrangian-vertices},
\\
R^{\text{lat},(0)}_{\text{HQ}}(p+q,p) &=
R^{(0)}_{\text{HQ}}(p+q,p),
\qquad
R^{\text{lat},(1)}_{\text{HQ},\mu}(p+q,p) =
R^{(1)}_{\text{HQ},\mu}(p+q,p) .
\label{eq:matching-current-HQET}
\end{align}
From the explicit expressions for
$\Lambda^{\text{(lat)},(0)}_{\text{HQ}}$ and
$\Lambda^{\text{(lat)},(1)}_{\text{HQ},\mu}$, one can show that
Eqs.~\eqref{eq:match-lagrangian-vertices} are satisfied automatically
from matching the OK action \cite{PhysRevD.78.014504}.
This consistency is highly non-trivial.
The OK action given in Ref.~\cite{PhysRevD.78.014504} passes multiple
consistency tests up to $\mathcal{O}(\lambda^3)$.
By imposing Eqs.~\eqref{eq:matching-current-HQET} at each order
through $\mathcal{O}(\lambda^3)$, we obtain a complete set of
constraints to determine all the improvement parameters $d_i$ in
Eq.~\eqref{eq:improved-current-lambda-third}.


\section{Results}
\label{sec:results}

In summary, by matching the four-quark matrix elements of Eq.~\eqref{eq:4q-ME},
the current operator defined by the improved quark field of
Eq.~\eqref{eq:improved-current-lambda-third} is matched to the lattice HQET
current:
\begin{align}
\label{eq:HQET-lattice-FWT}
\bar{\Psi}_{Ic}\Gamma \Psi(x)_{Ib}
\doteq \bar{h}_c \bar{\mathscr{U}}^c_{\text{lat}} \Gamma \mathscr{U}^b_{\text{lat}} h_b,
\end{align}
where
\begin{align}
&\mathscr{U}^b_{\text{lat}} =
\Bigg[ 1- \frac{1}{2m_{3,b}} \boldsymbol{\gamma \cdot D}
+\frac{1}{8m^2_{D_\perp^2,b}}\boldsymbol{D}^2
+\frac{i}{8m^2_{sB,b}}\boldsymbol{\Sigma}\cdot \boldsymbol{B}
+\frac{1}{4m^2_{\alpha E, b}}\boldsymbol{\alpha}\cdot \boldsymbol{E}
-\frac{\{\gamma_4D_4,\boldsymbol{\alpha}\cdot \boldsymbol{E}\}}{8m^3_{\alpha EE,b}}
-\frac{3\{\boldsymbol{\gamma \cdot D},\boldsymbol{D}^2\}}{32m_{\gamma D D_\perp^2,b}^3}
\nonumber \\
&
-\frac{3\{\boldsymbol{\gamma \cdot D},i\boldsymbol{\Sigma}\cdot \boldsymbol{B}\}}{32m_{5,b}^3}
-\frac{\{\boldsymbol{\gamma}\cdot \boldsymbol{D},
\boldsymbol{\alpha}\cdot \boldsymbol{E}\}}{16m_{\alpha_{rE},b}^3}
+\frac{[\gamma_4 D_4,\boldsymbol{D}^2]}{16m_{6,b}^3}
+\frac{[\gamma_4 D_4,i\boldsymbol{\Sigma} \cdot \boldsymbol{B}]}{16m_{7,b}^3}
+dw_{1,b} \sum_i \gamma_i D_i^3
+\frac{dw_{2,b}}{8} [\boldsymbol{\gamma}\cdot \boldsymbol{D},\boldsymbol{D}^2]
\Bigg], \label{eq:hqet-current-ops}
\end{align}
and $dw_{i,b} = 0$ and $m_{i,b} = m_b$ are required for matching.  
Explicit expressions for $dw_{i,b}$ and $m_{i,b}$ are given in
Appendix~\ref{app:coeff}.
The operators in Eq.~\eqref{eq:hqet-current-ops} are the continuum operators of
the HQET description of lattice QCD, and the coefficients contain the lattice
artifacts of the heavy quark.
Although the matching calculation \textit{via} HQET is distinct from
our previous calculations~\cite{Bailey:2016wza}, the results for all
improvement parameters except $d_{r_E}$ (and the new coefficients
$d_6$ and $d_7$) are the same.

The results for the improvement parameters are as follows.
Two of the improvement parameters are zero:  $d_{z_3}=d_{z_E}=0$.
We omit the flavor index ($b$ or $c$) for notational convenience.
The remaining results are
\begin{align}
d_1 & = \frac{\zeta(1+m_0)}{m_0(2+m_0)}-\frac{1}{2m},
\qquad
d_2  = 
\frac{2\zeta(1+m_0)}{m_0(2+m_0)}d_1
-\frac{r_s\zeta}{2(1+m_0)}
- \frac{\zeta^2(1+m_0)^2}{m_0^2(2+m_0)^2}
+\frac{1}{4m^2},
\\
d_3 &= w_3 -d_1,
\qquad
d_B  = 
\frac{2\zeta(1+m_0)}{m_0(2+m_0)}d_1
-\frac{c_B\zeta}{2(1+m_0)}
- \frac{\zeta^2(1+m_0)^2}{m_0^2(2+m_0)^2}
+\frac{1}{4m^2},
\\
d_E &= 
-\frac{2(1+m_0)\zeta}{m_0^2(2+m_0)^2}-\frac{(m_0+1)\zeta c_E}{m_0(2+m_0)}
+\frac{1}{2m^2},
\qquad
d_{r_E}  = \frac{1}{16m_3m^2_{\alpha E}} +\frac{d_1d_E}{4} - \frac{1}{16m^3},
\\
d_{EE} &= \frac{1+m_0}{(m_0^2+2m_0+2)}\left[-\frac{1}{4m^3}
+\frac{\zeta(1+m_0)(m_0^2+2m_0+2)}{[m_0(2+m_0)]^3}
+\frac{\zeta c_E(1+m_0)}{[m_0(2+m_0)]^2}
+\frac{(2+2m_0+m_0^2)c_{EE}}{m_0(2+m_0)}\right],
\\
d_4 &= 
\frac{\zeta^3(m_0^3+3m_0^2+5m_0+3)}{2m_0^3(2+m_0)^3}
+\frac{r_s\zeta^2(3m_0^2+6m_0+4)}{4m_0^2(2+m_0)^2}
+\frac{2(1+m_0)c_2}{m_0(2+m_0)}
-\frac{(1+m_0)^2\zeta^2}{2m_0^2(2+m_0)^2}d_1
\nonumber \\
&
-\frac{r_s\zeta}{4(1+m_0)}d_1
+\frac{(1+m_0)\zeta d_2}{2m_0(2+m_0)}-\frac{3}{16m^3},
\\
d_5 &= \frac{1}{2}\Bigg[ 
\frac{\zeta^3(m_0^3+3m_0^2+5m_0+3)}{2m_0^3(2+m_0)^3}
+\frac{c_B\zeta^2(3m_0^2+6m_0+4)}{4m_0^2(2+m_0)^2}
+\frac{2(1+m_0)c_3}{m_0(2+m_0)}
-\frac{(1+m_0)^2\zeta^2}{2m_0^2(2+m_0)^2}d_1
\nonumber \\
&
-\frac{c_B\zeta}{4(1+m_0)}d_1
+\frac{(1+m_0)\zeta d_B}{2m_0(2+m_0)}-\frac{3}{16m^3}\Bigg],
\\
d_6 & =
\frac{2(1+m_0)}{(m_0^2+2m_0+2)}\Bigg[ - \frac{1}{16m_3m^2_{\alpha E}} 
+\frac{\zeta^2 c_E}{4m_0(2+m_0)} 
-\frac{\zeta c_{EE}(m_0^2+2m_0+2)}{2m_0(1+m_0)(2+m_0)}
-\frac{d_E}{4}\Big(d_1-\frac{2\zeta(1+m_0)}{m_0(2+m_0)} \Big)
\nonumber \\
&-\frac{1}{24m_2} +\frac{1}{16m^3}\Bigg],
\\
d_7 & =
\frac{2(1+m_0)}{(m_0^2+2m_0+2)}\Bigg[ - \frac{1}{16m_3m^2_{\alpha E}} 
+\frac{\zeta^2 c_E}{4m_0(2+m_0)} 
-\frac{\zeta c_{EE}(m_0^2+2m_0+2)}{2m_0(1+m_0)(2+m_0)}
-\frac{d_E}{4}\Big(d_1-\frac{2\zeta(1+m_0)}{m_0(2+m_0)} \Big)
\nonumber \\
&-\frac{1}{24m_B} +\frac{1}{16m^3}\Bigg].
\label{eq:result_end}
\end{align}

The parameters $d_1$, $d_2$, $d_3$, and $d_4$ are determined by
matching the zero-gluon (current) vertex.
The remaining parameters are determined by matching the one-gluon
(current) vertex.
The result for $d_5$ appears very different from that obtained from
our previous matching calculation~\cite{Bailey:2016wza}.
In fact the results are identical after matching.
The result for $d_{r_E}$ is modified by the addition of the $d_6$
operator to the improved quark field, which makes our previous
calculation of $d_{r_E}$ \cite{Bailey:2016wza} obsolete.
As noted in Ref.~\cite{Bailey:2016wza}, the result for $d_E$ is
different from that given in Ref.~\cite{ElKhadra:1996mp}.
We anticipate that matching the four-quark matrix elements
\textit{via} NRQCD, in analogy with the matching calculation
\textit{via} HQET reported here, will yield the same results as in
Ref.~\cite{ElKhadra:1996mp}.

Finally, we note that the matching conditions yield about 150
constraints for the 13 improvement parameters $d_i$.
These constraints also involve the coefficients in the improvement
terms of the OK action.
The results of the matching calculation reported here, together with
the OK-action coefficients, are consistent with all the constraints.
The results of this paper are being used to calculate semileptonic
form factors for $\bar{B}\to D^\ast \ell \bar{\nu}$ decay \cite{
  SW:2017proc}.
Meanwhile, additional cross checks are underway to confirm the
results.
%


\section*{Acknowledgements}
\begin{acknowledgement}
  The research of W.~Lee is supported by the Creative Research
  Initiatives Program (No.~2017013332) of the NRF grant funded by the
  Korean government (MEST).
  W.~Lee would like to acknowledge the support from the KISTI
  supercomputing center through the strategic support program for the
  supercomputing application research (No.~KSC-2015-G2-002).
  Computations were carried out on the DAVID clusters at Seoul
  National University.
  J.A.B. is supported by the Basic Science Research Program of the
  National Research Foundation of Korea (NRF) funded by the Ministry of
  Education (No.~2015024974).
\end{acknowledgement}

\appendix
\section{\label{app:lhqet}Lattice HQET vertices}

In this and the following section, we omit the flavor index ($b$ or $c$) for 
convenience.
The zero-gluon vertex of the lattice HQET Lagrangian is given by
\begin{align}
\label{eq:lat-HQET-zero-gluon-lagrangian}
\Lambda^{\text{lat},(0)}_{\text{HQ}}(p) = -\frac{1}{2m_2}\boldsymbol{p}^2
+\frac{1}{8m_4^3}\big(\boldsymbol{p}^2\big)^2
+\frac{1}{6}w_4 \sum_i p_i^4,
\end{align}
and the one-gluon vertex is 
$(-gt^a)\Lambda^{\text{lat},(1)}_{\text{HQ},\mu}(p+q,p)$, where 
\begin{align}
&\Lambda^{\text{lat},(1)}_{\text{HQ},4}(p+q,p)
=\Big[
 1- \frac{\boldsymbol{q}^2 - 2i \epsilon_{ijk} q_i p_j \Sigma_k}{8m_E^2} \Big],
\\
&\Lambda^{\text{lat},(1)}_{\text{HQ},i}(p+q,p)
= \Big[
-\frac{i}{2m_2}(2p_i+q_i)
+\frac{1}{2m_B}\epsilon_{ijk} \Sigma_j q_k
+\frac{q_4}{8m_E^2}\big(q_i+ i\epsilon_{ijk}\Sigma_j (2p_k+q_k)\big)
\nonumber \\
&
+\frac{i(2p_i+q_i)}{8m_4^3}
\big((\boldsymbol{p}+\boldsymbol{q})^2+\boldsymbol{p}^2\big)
-\frac{1}{8m_{B^{\prime}}^3}\epsilon_{ijk}\Sigma_j q_k 
\big((\boldsymbol{p}+\boldsymbol{q})^2+\boldsymbol{p}^2\big)
+\frac{i}{6}w_4 (2p_i +q_i)\big((p_i+q_i)^2+p_i^2\big)
\nonumber \\
&
-\frac{i}{8}w_{B}\big(p_i \boldsymbol{q}^2 - q_i \boldsymbol{p}\cdot
\boldsymbol{q}\big)
-\frac{1}{16}w_{B}\epsilon_{ijk}\Sigma_j q_k \boldsymbol{q}^2
-\frac{1}{8}w_{B}\epsilon_{ijk}q_j p_k \boldsymbol{\Sigma}\cdot(2\boldsymbol{p}+\boldsymbol{q})
-\frac{1}{12}w^\prime_B \epsilon_{ijk}\Sigma_j q_k (q_i^2 + q_k^2) 
\nonumber \\
&
-\frac{1}{12}(w_4+w^\prime_4)\epsilon_{ijk} \Sigma_j q_k \Big((3p_i^2 + 3p_i q_i + q_i^2)
+ (3p_k^2 + 3p_k q_k+ q_k^2) \Big)
\Big].
\label{eq:lat-HQET-one-gluon-lagrangian}
\end{align}
The zero-gluon vertex of the lattice HQET flavor-changing current is
%
%
\begin{align}
\label{eq:lat-HQET-zero-gluon-operator}
R^{\text{lat},(0)}_{\text{HQ}}(p) = 
1&-\frac{i}{2m_3}\boldsymbol{\gamma}\cdot \boldsymbol{p}
-\frac{1}{8m^2_{D^2_\perp}}\boldsymbol{p}^2
+\frac{3i\boldsymbol{\gamma}\cdot \boldsymbol{p}}{16m^3_{D^3_\perp}}\boldsymbol{p}^2
-dw_1 \sum_j i\gamma_j p^3_j,
\end{align}
and the one-gluon vertex 
is $(-gt^a)R^{\text{lat},(1)}_{\text{HQ},\mu}(p+q,p)$, with
\begin{align}
&R^{\text{lat},(1)}_{\text{HQ},4}(p+q,p) 
= 
-\frac{i\boldsymbol{\gamma}\cdot \boldsymbol{q}}{4m_{\alpha E}^2}
+\frac{q_4\boldsymbol{\gamma}\cdot \boldsymbol{q}}{8m_{\alpha {EE}}^3}
-\frac{\big(\boldsymbol{q}^2 -2i\epsilon_{ijk}\Sigma_i q_jp_k\big)}{16m_{\alpha_{rE}}^3}
-\frac{\big(\boldsymbol{q}^2 + 2\boldsymbol{p}\cdot \boldsymbol{q} \big)}{16m_6^3},
\\
&R^{(1)}_{\text{HQ},i}(p+q,p) = 
\frac{1}{2m_3}\gamma_i 
+\frac{i q_4}{4m_{\alpha E}^2}\gamma_i
-\frac{i}{8m_{D^2_\perp}^2}(2p_i+q_i) 
+\frac{\epsilon_{ijk}\Sigma_j q_k}{8m_{sB}^2}
-\frac{q_4^2}{8m_{\alpha {EE}}^3}\gamma_i 
\nonumber \\
&-\frac{3}{32m_{D^3_\perp}^3}\big(
\boldsymbol{\gamma}\cdot (2\boldsymbol{p}+\boldsymbol{q})(2p_i+q_i) 
+(\boldsymbol{p}^2+(\boldsymbol{p}+\boldsymbol{q})^2)\gamma_i\big)
-\frac{3i\epsilon_{ijk}q_k }{32m^3_{5}}
\big(\Sigma_j \boldsymbol{\gamma}\cdot\boldsymbol{p}
+\boldsymbol{\gamma}\cdot(\boldsymbol{p}+\boldsymbol{q})\Sigma_j \big)
\nonumber \\
&
+\frac{q_4}{16m_{\alpha_{rE}}^3}\big(i\epsilon_{ijk}\Sigma_j(2p_k+q_k) +q_i\big)
+\frac{q_4}{16m_6^3}(2p_i+q_i)
+\frac{q_4}{16m_7^3}i\epsilon_{ijk}\Sigma_j q_k
\nonumber \\
&
+dw_1 \gamma_i(3p_i^2 +3p_i q_i+q_i^2)
+\frac{dw_2}{8} \big(\boldsymbol{q}\cdot (2\boldsymbol{p}+\boldsymbol{q}) \gamma_i
+\boldsymbol{\gamma}\cdot \boldsymbol{q}(2p_i+q_i)\big).
\label{eq:lat-HQET-one-gluon-operator}
\end{align}

\section{\label{app:coeff}Short-distance coefficients}

The explicit formulas for the lattice mass parameters $m_i$ and the parameters
$w_{i}$ which appear in
\eqref{eq:lat-HQET-zero-gluon-lagrangian}-\eqref{eq:lat-HQET-one-gluon-lagrangian}
are given in Ref.~\cite{PhysRevD.78.014504}.
The other mass parameters and parameters $dw_i$ which appear in
\eqref{eq:lat-HQET-zero-gluon-operator}-\eqref{eq:lat-HQET-one-gluon-operator}
are as follows.
\begin{align}
\frac{1}{2m_3} &=  \frac{\zeta(1+m_0)}{m_0(2+m_0)}-d_1,
\qquad
\frac{1}{4m^2_{\alpha E}} = 
\frac{(1+m_0)\zeta}{m_0^2(2+m_0)^2}+\frac{(1+m_0)\zeta c_E}{2m_0(2+m_0)}
+\frac{d_E}{2},
\\
\frac{1}{8m^2_{D_\perp^2}} 
&=  
-\frac{\zeta(1+m_0)}{m_0(2+m_0)}d_1
+\frac{r_s\zeta}{4(1+m_0)}
+ \frac{\zeta^2(1+m_0)^2}{2m_0^2(2+m_0)^2}
+\frac{d_2}{2},
\\
\frac{1}{8m^2_{sB}} 
&= 
-\frac{\zeta(1+m_0)}{m_0(2+m_0)}d_1
+\frac{c_B\zeta}{4(1+m_0)}
+ \frac{\zeta^2(1+m_0)^2}{2m_0^2(2+m_0)^2}
+\frac{d_B}{2},
\\
\frac{1}{16m_{\alpha_{rE}}^3} &= \frac{1}{16m_3m^2_{\alpha E}} +\frac{d_1d_E}{4} -d_{r_E},
\\
\frac{1}{16m_{\alpha {EE}}^3} &= \frac{(1+m_0)(m_0^2+2m_0+2)\zeta}{4m_0^3(2+m_0)^3}
+\frac{(1+m_0)\zeta c_E}{4m_0^2(2+m_0)^2}
+\frac{(m_0^2+2m_0+2)c_{EE}}{4m_0(2+m_0)}-\frac{(m_0^2+2m_0+2)d_{EE}}{4(1+m_0)},
\\
\frac{3}{16m_{\gamma D D_\perp^2}^3} &= 
\frac{\zeta^3(m_0^3+3m_0^2+5m_0+3)}{2m_0^3(2+m_0)^3}
+\frac{r_s\zeta^2(3m_0^2+6m_0+4)}{4m_0^2(2+m_0)^2}
+\frac{2(1+m_0)c_2}{m_0(2+m_0)}
\nonumber \\
&-\frac{(1+m_0)^2\zeta^2}{2m_0^2(2+m_0)^2}d_1
-\frac{r_s\zeta}{4(1+m_0)}d_1
+\frac{(1+m_0)\zeta d_2}{2m_0(2+m_0)}-d_4,
\\
\frac{3}{16m_{5}^3} &= 
\frac{\zeta^3(m_0^3+3m_0^2+5m_0+3)}{2m_0^3(2+m_0)^3}
+\frac{c_B\zeta^2(3m_0^2+6m_0+4)}{4m_0^2(2+m_0)^2}
+\frac{2(1+m_0)c_3}{m_0(2+m_0)}
\nonumber \\
&-\frac{(1+m_0)^2\zeta^2}{2m_0^2(2+m_0)^2}d_1
-\frac{c_B\zeta}{4(1+m_0)}d_1
+\frac{(1+m_0)\zeta d_B}{2m_0(2+m_0)}-2d_5,
\\
\frac{1}{16m_{6}^3} &= \frac{1}{16m_3m^2_{\alpha E}} -\frac{\zeta^2 c_E}{4m_0(2+m_0)} 
+\frac{\zeta c_{EE}(m_0^2+2m_0+2)}{2m_0(1+m_0)(2+m_0)}
+\frac{d_E}{4}\Big(d_1-\frac{2\zeta(1+m_0)}{m_0(2+m_0)} \Big)
+\frac{1}{24m_2}
\nonumber \\
&+\frac{(m_0^2+2m_0+2)}{2(1+m_0)}d_6,
\\
\frac{1}{16m_{7}^3} &= \frac{1}{16m_3m^2_{\alpha E}} -\frac{\zeta^2 c_E}{4m_0(2+m_0)} 
+\frac{\zeta c_{EE}(m_0^2+2m_0+2)}{2m_0(1+m_0)(2+m_0)}
+\frac{d_E}{4}\Big(d_1-\frac{2\zeta(1+m_0)}{m_0(2+m_0)} \Big)
+\frac{1}{24m_B}
\nonumber \\
&+\frac{(m_0^2+2m_0+2)}{2(1+m_0)}d_7,
\\
dw_1 &= d_3 +d_1 -w_3, 
\qquad
dw_2 = 
\frac{\zeta^2(r_s-c_B)+2\zeta (d_2-d_B)(1+m_0)}{m_0(2+m_0)}.
\label{eq:short-distance-coeff}
\end{align}


\bibliography{ref}

\end{document}